\documentclass{Interspeech}

\interspeechcameraready

\title{Differentiable K-means for Fully-optimized Discrete Token-based ASR}
\author[affiliation={1,2}]{Kentaro}{Onda}
\author[affiliation={2}]{Yosuke}{Kashiwagi}
\author[affiliation={2}]{Emiru}{Tsunoo}
\author[affiliation={2}]{Hayato}{Futami}
\author[affiliation={3}]{Shinji}{Watanabe}

\affiliation{}{The University of Tokyo}{Japan}
\affiliation{}{Sony Group Corporation}{Japan}
\affiliation{}{Carnegie Mellon University}{USA}
\email{ondakentaro@gavo.t.u-tokyo.ac.jp}

\usepackage{comment}
\usepackage{multirow}
\usepackage{array}
\usepackage{cite}

\newcommand{\argmax}{\mathop{\rm arg~max}\limits}
\newcommand{\argmin}{\mathop{\rm arg~min}\limits}

\begin{document}

\maketitle

\begin{abstract}
   Recent studies have highlighted the potential of discrete tokens derived from self-supervised learning (SSL) models for various speech-related tasks. These tokens serve not only as substitutes for text in language modeling but also as intermediate representations for tasks such as automatic speech recognition (ASR). However, discrete tokens are typically obtained via k-means clustering of SSL features independently of downstream tasks, making them suboptimal for specific applications. This paper proposes the use of differentiable k-means, enabling the joint optimization of tokenization and downstream tasks. This approach enables the fine-tuning of the SSL parameters and learning weights for outputs from multiple SSL layers. Experiments were conducted with ASR as a downstream task. ASR accuracy successfully improved owing to the optimized tokens. The acquired tokens also exhibited greater purity of phonetic information, which were found to be useful even in speech resynthesis. 
\end{abstract}

\section{Introduction}
The rapid advancement of Self-Supervised Learning (SSL) techniques has significantly influenced numerous speech processing tasks \cite{sslreview}. 
Pioneering SSL-based models such as HuBERT \cite{hubert}, wav2vec 2.0 \cite{wav2vec}, WavLM \cite{Chen2021WavLMLS}, and w2v-BERT \cite{Chung2021w2vBERTCC} have demonstrated strong performance across diverse applications including automatic speech recognition (ASR), emotion analysis, and text-to-speech (TTS) synthesis \cite{yang21c_interspeech}. 
One notable feature of these models is their ability to learn robust and generalizable representations of audio signals without extensive labeled data. 
Moreover, joint optimization strategies, in which the SSL model is fine-tuned on downstream objectives, have consistently yielded further performance gains. 
This underscores the critical importance of end-to-end training in speech-related tasks using SSL models \cite{hubert, Chen2021WavLMLS}.

While continuous SSL embeddings are already widely used, there is a growing tendency to discretize these learned representations \cite{chang24b_interspeech}. 
Typically obtained via methods such as k-means clustering, these discrete tokens have proven highly effective for tasks that benefit from a text-like representation. 
They are also integrated seamlessly with large language models (LLMs), opening new possibilities in speech and language research, notably in generative frameworks such as SpeechGPT \cite{zhang2023speechgpt}, VoxtLM \cite{maiti2024voxtlm}, and AudioPaLM \cite{rubenstein2023audiopalm}.

A major limitation of relying on k-means is its non-differentiability. 
The conventional SSL pipeline commonly involves three steps: (1) SSL pre-training, (2) k-means clustering to obtain tokens, and (3) downstream training. 
Such a procedure restricts the ability to fine-tune the SSL model's parameters to the final task objectives \cite{chang23b_interspeech, chang2024exploring,yang2024towards}. 
Consequently, the resulting tokens may not capture task-relevant information optimally, leading to a performance gap compared with systems that leverage continuous SSL features and end-to-end optimization \cite{yeh_2024slt}.

To address this shortcoming, we propose a framework that integrates differentiable k-means into a discrete token-based ASR system, thereby enabling joint optimization of both the SSL model and the clustering process. 
Our approach is inspired by previous work on differentiable k-means in autoencoders \cite{Gao2020}, but extends this idea to SSL-based speech representations. 
By making the tokenization stage differentiable, we can backpropagate error signals from the downstream ASR objective into the SSL feature extractor. 
While discretization techniques such as those in VQ-VAE \cite{van2017neural} could be employed for this purpose, we adopt differentiable k-means as a more straightforward extension of standard k-means, which is widely used for tokenizing SSL representations.
Moreover, the framework supports simultaneous optimization of layer-weighting factors when multiple SSL layers are used \cite{yang2024towards}, which further refines the quality of the learned discrete tokens. 
This end-to-end design also offers greater flexibility in how clustering boundaries are formed, allowing the model to focus on phonetic distinctions that are most relevant for the final recognition task. 
As a result, the system effectively learns discrete tokens that discard extraneous speaker or noise information.
Our experimental results indicate that this method narrows the performance gap relative to systems using continuous SSL features in ASR while also providing an efficient representation that benefits both ASR and generative tasks.

Our key contributions can be summarized as follows: 
\begin{itemize} 
    \item \textbf{Differentiable Tokenization for ASR:} We illustrate how integrating differentiable k-means into a discrete token-based ASR system enables end-to-end training, allowing each component (from SSL parameters to clustering centroids) to be jointly optimized for the recognition task. 
    \item \textbf{Enhanced Representational Flexibility:} We demonstrate that our method allows layer-weighting strategies for exploiting multiple SSL layers, thereby enabling more flexible and effective use of the hierarchical representations learned during SSL pre-training. 
    \item \textbf{Analysis of Updated Tokenization:} Through an analytical investigation, we show that the discrete tokens learned under an ASR-centric objective become increasingly similar to phoneme-level representations. 
    This property not only improves ASR accuracy but also holds potential for generative applications. 
\end{itemize}


\section{Discrete Token-based ASR with Differentiable K-means}

\subsection{Discrete token-based ASR}
Discrete token-based ASR has recently attracted attention as an alternative to using continuous features extracted from SSL models \cite{chang23b_interspeech, chang2024exploring, yang2024towards, mousavi24_interspeech, shi24h_interspeech}. 
In this paradigm, each frame of the SSL output is clustered into discrete tokens, which then serve as input to a conventional sequence-to-sequence (Seq2Seq) ASR model or a similar architecture \cite{ctcaed, graves2012sequence, peng24b_interspeech}. 
Let \(X\) denote the input speech and \(Y\) denote the corresponding text token sequence. 
We denote the intermediate discrete token sequence by \(Z\), and the ASR objective is typically formulated as minimizing a cross-entropy loss \(\mathcal{L}^{\text{asr}}(Y, \mathrm{ASR}(Z; \theta_{\text{asr}}))\), where \(\theta_{\text{asr}}\) represents ASR model parameters.

Although continuous SSL features often yield higher recognition accuracy, discrete tokens offer several advantages. 
They can represent speech frames in a compact manner by clustering similar frames and possibly applying deduplication, which shortens the input sequence for the subsequent model. 
In addition, discrete tokens approximate a “text-like” representation that is highly compatible with LLMs. 
This property facilitates downstream tasks such as ASR and speech-to-text translation, where LLMs can be leveraged or fine-tuned from large-scale text-based pretraining \cite{zhang2023speechgpt, maiti2024voxtlm, rubenstein2023audiopalm, xu24d_interspeech}.
One of the main challenges, however, is that the discrete clustering step in most existing pipelines relies on k-means, a non-differentiable algorithm that prevents error backpropagation from the final ASR objective into the SSL parameters.

\subsection{Differentiable k-means}
Standard k-means clustering assigns each data point \(\vec{s_i}\), 
the output from the SSL model,
to the nearest centroid \(\vec{\mu_j}\) based on Euclidean distance. 
Since this involves a discrete selection (\(\argmin\)), it is inherently non-differentiable. 
To make k-means amenable to gradient-based optimization, \cite{Gao2020} proposed an approach that uses the Gumbel-Softmax reparameterization trick \cite{jang2017categorical}. 
Concretely, the probability of assigning \(\vec{s_i}\) to cluster $j$
is defined as:
\begin{align}
    p(j \mid \vec{s_i})
    = 
    \frac{\exp\bigl(-\sigma^2 \lVert \vec{s_i} - \vec{\mu_j} \rVert^2_2\bigr)}
         {\sum_{c=1}^{k} \exp\bigl(-\sigma^2 \lVert \vec{s_i} - \vec{\mu_c}\rVert^2_2\bigr)},
    \label{eq:kmeans_prob}
\end{align}
where \(\sigma\) is a hyperparameter (or learnable parameter), and \(k\) is the number of clusters. 
This distribution \(p(j \mid \vec{s_i})\) is then sampled via Gumbel-Softmax:
\begin{align}
    h_{i,j}
    =
    \frac{\exp\bigl(\tau^{-1}(\log p(j\mid \vec{s_i}) + G_j)\bigr)}
         {\sum_{c=1}^{k} \exp\bigl(\tau^{-1}(\log p(c\mid \vec{s_i}) + G_c)\bigr)},
    \label{eq:gumbel}
\end{align}
where \(G\) is noise drawn from the Gumbel distribution, and \(\tau\) is a temperature parameter. 
As \(\tau \to 0\), \(\vec{h}_{i} = [h_{i,1}, ..., h_{i,k}]\) approaches a one-hot vector. 
Although Gumbel-Softmax produces discrete outputs, i.e. $\argmax h_i$, it internally handles assignments in a probabilistic manner.
Therefore it enables backpropagation to both the output from SSL model \(\vec{s}_i\) and the centroids \(M = \{\vec{\mu_1}, \ldots, \vec{\mu_k}\}\).
The k-means loss for optimizing the centroids \(M\) is given by:
\begin{align}
    \mathcal{L}^{\text{km}} (S, M)
    =
    \sum_{i=1}^{N}
    \Bigl\lVert\,
        \vec{s_i} - \tilde{\vec{h}}_i \, M
    \Bigr\rVert_2^2,
    \label{eq:kmeans_loss}
\end{align}
where N is the total number of data points, and
\(\tilde{\vec{h}}_i\) is the discrete one-hot version of \(\vec{h}_{i}\). $S$ is a set of $\vec{s_i}$.
Differentiable k-means was originally developed for autoencoders, but can be applied in a more general context by substituting the relevant downstream loss.

\subsection{Error Backpropagation via differentiable k-means}
We extend this differentiable k-means approach to discrete token-based ASR. 
Unlike conventional pipelines, which cluster SSL features into discrete tokens in a downstream-agnostic manner, our method enables joint optimization of clustering and recognition. 
Let \(\theta_{\text{ssl}}\) be the parameters of the SSL model and \(\theta_{\text{asr}}\) be those of the ASR model. 
The overall loss is:
\begin{align}
\label{eq:joint_loss}
\mathcal{L}
&=
\mathcal{L}^{\text{asr}}\Bigl(
    Y,\,
    \mathrm{ASR}\bigl(
        \mathrm{DiffKM}\bigl(\mathrm{SSL}(X;\,\theta_{\text{ssl}}),\,M\bigr);\,
        \theta_{\text{asr}}
    \bigr)
\Bigr)
\nonumber\\
&\quad + \alpha\, \mathcal{L}^{\text{km}} \bigl(\mathrm{SSL}(X;\,\theta_{\text{ssl}}),\,M\bigr),
\end{align}
where \(\mathrm{DiffKM}(\cdot)\) denotes the differentiable k-means operator in Eqs.\,(\ref{eq:kmeans_prob})--(\ref{eq:gumbel}), 
and \(\alpha\) is a scalar that balances the cross-entropy loss \(\mathcal{L}^{\text{asr}}\) and the k-means loss \(\mathcal{L}^{\text{km}}\). 
By backpropagating through both cluster assignments and SSL feature extractor, the discrete tokens become optimized with respect to the ASR objective. 
In practice, \(\alpha\) can be tuned or set to zero if the ASR objective alone sufficiently drives discriminative clustering.
{\color{red}}

\subsection{Joint optimization with multilayer weighting}
\label{sec:joint_ws}
Combining outputs from multiple SSL layers often leads to performance gains, particularly when working with continuous representations \cite{yang21c_interspeech, mousavi24_interspeech, shi24h_interspeech}. 
In that continuous domain, a widely adopted approach is to take a weighted sum of each layer’s output before feeding it into a downstream model:
\begin{equation}
    \label{eq:weighted_sum}
    \mathrm{SSL}_{\text{ws}}(X;\theta_{\text{ssl}}, \vec{w})
    =
    \sum_{\ell=1}^{L} w_{\ell}\,\mathrm{SSL}_{\ell}(X;\theta_{\text{ssl}}),
\end{equation}
where \(L\) is the total number of SSL layers, \(\mathrm{SSL}_{\ell}(\cdot)\) is the output of the \(\ell\)th layer, and \(\vec{w}=[w_{1}, \dots, w_{L}]\) denotes the set of learnable layer weights.

Although extending this idea to discrete representations has been more challenging, our differentiable k-means-based framework resolves this issue. 
By substituting Equation \eqref{eq:weighted_sum} into Equation \eqref{eq:joint_loss}, we can obtain weighted-sum based differentiable loss function as:
\begin{align}
\label{eq:joint_loss_ws}
\mathcal{L}^{\text{asr}}\Bigl(
    Y,\,
    \mathrm{ASR}\bigl(
        \mathrm{DiffKM}\bigl(\mathrm{SSL}_{\text{ws}}(X;\theta_{\text{ssl}}, \vec{w}), M\bigr);\,
        \theta_{\text{asr}}
    \bigr)
\Bigr)
\end{align}
Thus, we can jointly optimize \(\vec{w}\), the cluster centroids \(M\), and even the SSL model parameters \(\theta_{\text{ssl}}\) for the final ASR objective. 
This design ensures a single token is generated for each frame of the fused representation, removing unnecessary redundancy. 
In addition, every component of the system can be trained cooperatively, which avoids ad hoc decisions about how to merge tokens from different layers and allows us to exploit the rich complementary information in a more principled manner.

\section{Experiments}
\subsection{Experimental Setup}
Our experiments were conducted using ESPnet \cite{watanabe18_interspeech}, following the basic configurations described in \cite{chang23b_interspeech, chang2024exploring}.
We used the joint CTC/attention-based encoder-decoder model \cite{ctcaed} with WavLM-large\footnote{\url{https://huggingface.co/microsoft/wavlm-large}} as the upstream SSL. 
To avoid additional complexity in handling gradients, we did not perform deduplication or subword modeling 
on the token sequences.
Regarding the hyperparameters, 
$\sigma^2$ in Equation \eqref{eq:kmeans_prob} was fixed at 1.0. $\tau$ in Equation \eqref{eq:gumbel} was initialized to 2.0
and gradually decreased during training.
$\alpha$ in Equation \eqref{eq:joint_loss} was set to 0, as a smaller value led to better ASR performance in our preliminary experiments.
During decoding, the beam size was set to 5,
which may result in slightly lower performance compared to other studies. 
However, the same settings were applied in all experiments to ensure a consistent evaluation of the proposed method's effectiveness.

\subsection{Single-layer settings with LibriSpeech100}
\label{subsec:libri100}

We first conducted experiments on LibriSpeech-100 \cite{libri} 
using the 21st layer of the SSL model by following \cite{chang2024exploring} for feature extraction.
We compared ASR performance with various cluster sizes.
In all cases, training was conducted for 90 epochs,
using the loss shown in Equation (\ref{eq:joint_loss}).
During the first 30 epochs, the SSL model ($\theta_{\text{ssl}}$) and centroids ($M$) were frozen, and only the 
downstream ASR module ($\theta_{\text{asr}}$) was trained.
We compared the following three conditions for the parameter updates applied during the remaining 60 epochs:
\textbf{baseline}) ASR model $\theta_{\text{asr}}$ only, 
\textbf{freeze SSL}) ASR model $\theta_{\text{asr}}$ and centroids $M$, and 
\textbf{full finetune}) ASR model $\theta_{\text{asr}}$, centroids $M$, and SSL model $\theta_{\text{ssl}}$.
In all cases, $M$ was initialized using standard k-means, pre-trained on the train\_clean\_100. 
Here, \textbf{baseline} functions equivalently to a conventional discrete token-based ASR.

Table \ref{tab:single} shows the results.
In all cases, the proposed method (\textbf{full finetune}) demonstrated the highest performance.
For cluster sizes, we obtained larger improvements, particularly in the small cluster size.
Generally, increasing the number of clusters leads to higher recognition accuracy.
However, with the proposed method, there was no significant difference
in performance between cluster sizes of 500, 1000, and 2000. 
This indicates that differentiable k-means effectively enabled
the representation of necessary linguistic information 
with a smaller number of units.
This characteristic
also offers an advantage from the perspective of data compression.
For \textbf{freeze SSL},
no significant improvement was observed.

\begin{table}[tb]
    \centering
    \caption{LibriSpeech-100 results with single-layer settings: WER(\%) of  test-\{clean, other\}}
    \label{tab:single}
    \vspace*{-2mm}
    \resizebox{0.8\columnwidth}{!}{
    \begin{tabular}{lccc}
    \toprule
        cluster size & \textbf{baseline} & \textbf{freeze SSL} & \textbf{full finetune} \\\midrule
        100 &7.6/13.2 &7.3/12.6 & \textbf{5.4/9.2}\\
        500 & 4.7/\phantom{1}7.9&4.8/\phantom{1}7.9 & \textbf{4.3/7.1}\\
        1000 & 4.3/\phantom{1}7.5& 4.2/\phantom{1}7.4 & \textbf{3.9/6.8}\\
        2000 &4.4/\phantom{1}7.2 & 4.4/\phantom{1}7.2 & \textbf{4.2/6.8}\\
        \bottomrule
    \end{tabular}

    }
     \vspace*{-2mm}
\end{table}

\subsection{Multi-layer settings with LibriSpeech100}
\label{subsec:multi}

\begin{table}[tb]
    \centering
    \caption{LibriSpeech-100 results with multi-layer settings: WER(\%) of test-\{clean, other\}}
    \label{tab:multi}
    \vspace*{-2mm}
    \resizebox{0.8\columnwidth}{!}{
    \begin{tabular}{lccc}
    \toprule
        cluster size & \textbf{baseline} & \textbf{freeze SSL} & \textbf{full finetune} \\\midrule
        100  &   8.4/18.3     &    7.5/15.1  &   \textbf{5.3/8.2}  \\
        2000 & 5.4/10.8 & 4.7/\phantom{1}8.8 & \textbf{4.2/6.8}\\\midrule\midrule
        continuous & 3.7/\phantom{1}6.7     & 3.6/\phantom{1}6.7          & 3.7/6.6 \\
        \bottomrule
    \end{tabular}
    }
     \vspace*{-4mm}
\end{table}

Next, we explored the cases where outputs from multiple layers of the SSL model are used
with a weighted summation, as discussed in Section~\ref{sec:joint_ws}.
In the experiments, the outputs of all layers are used.
The training was also conducted for 90 epochs using the loss of Equation (\ref{eq:joint_loss_ws}).
As in single-layer settings, for the first 30 epochs,
only $\theta_{\text{asr}}$ was updated, and other parameters, including the layer weights ($\vec{w}$)
were frozen.
Similar to the previous experiment, the update with the remaining 60 epochs was applied to \textbf{baseline}) $\theta_{\text{asr}}$, \ \textbf{freeze SSL}) $\theta_{\text{asr}}$, $\vec{w}$, and $M$, or
\textbf{full finetune}) $\theta_{\text{asr}}$, $\vec{w}$, $M$ and $\theta_{\text{ssl}}$.
In the case of \textbf{full finetune}, the update of $\theta_{\text{ssl}}$ 
was performed only during the last 30 epochs. 
This is because $\vec{w}$ was effectively tuned in this case,
resulting in better recognition accuracy
compared to when SSL was updated from the beginning.
\vec{w} was initialized such that all of its elements to be the same.
$M$ was initialized using centroids obtained by applying standard k-means to a weighted sum computed with the initialized $\vec{w}$.

The results are shown in Table \ref{tab:multi}.
Experiments were conducted with cluster sizes of 100 and 2000,
and as a top line, results with the continuous SSL features are also presented.
In the case of continuous features, 
all SSL layers were used as well, and
the same strategy for parameter updating
was applied.
Similar to the single layer case in Section~\ref{subsec:libri100}, for both of the discrete cases, \textbf{full finetune} achieved the best results.
However, compared to Table\ref{tab:single}, \textbf{baseline} and \textbf{freeeze SSL} performed worse than the single-layer results, while \textbf{full finetune} was comparable.
These results indicate that while multi-layer approaches effectively leverage frozen SSL features by aggregating information from various layers, full fine-tuning can condense this information into a single layer optimized for downstream ASR tasks.
Notably, \textbf{full finetune} results approached top-line performance. This is a highly positive outcome, as it preserves both high performance and a compact discrete token representation simultaneously, without heuristic layer selection.


\subsection{Evaluation for token properties}
\label{subsec:cluster}

To investigate the impact of the ASR-centric fine-tuning on the tokenization process,
we evaluated the clusters themselves. 
Clustering quality is commonly evaluated  
based on their relationship with phonemes using metrics such as phone normalized mutual information (PNMI) \cite{hubert,lakhotiaetal2021generative,chang23b_interspeech}. 
In addition to this,
we explore three more evaluation metrics to examine the effect of fine-tuning more directly.

\noindent
\textbf{1) Normalized Quantization Error (NQE):} The average Euclidean distance between the continuous SSL features of each frame and the corresponding cluster centroid. 
Since SSL feature spaces are changed by finetuning,
they are divided by the averaged L2 norm for normalization.
As we use the ASR objective (Equation \eqref{eq:joint_loss} or \eqref{eq:joint_loss_ws}), which is different from the k-means objective (Equation \eqref{eq:kmeans_loss}), this value is expected to be higher with our proposed methods.

\noindent
\textbf{2) Token Sequence Length (TSL):} The length of token sequences after deduplication.
If the tokens are robust to subtle pronunciation variations and represent more phoneme-like symbols, 
the sequence increases the repetition and TSL will be shorter. 
A lower value also indicates a higher efficiency in data compression.

\noindent
\textbf{3) Mean Token Error Rate (MTER):} The edit distance between discrete token sequences from different utterances with the same transcriptions. 
A lower value indicates the robustness of discrete tokens 
to non-linguistic information. 
Unlike word error rate (WER) or character error rate (CER), 
which are calculated with the ground truth transcriptions,
there is no definitive correct token sequence in this case. Therefore, we prepare multiple utterances with the same transcriptions and calculate
token error rate (TER) for all possible pairs, then use the mean value for the metric.

\begin{table}[tb]
    \centering
    \caption{Cluster evaluations using TIMIT corpus}
    \label{tab:cluster}
   \vspace*{-2mm}
    \resizebox{1.0\columnwidth}{!}{
    \begin{tabular}{llc!{\vrule}c!{\vrule}cc}
    \toprule
        cluster& & PNMI &  NQE  & TSL& MTER [\%]\\        size& & ($\uparrow$) &   & ($\downarrow$)&  ($\downarrow$)\\\midrule
        100 & baseline &   0.4754 &  0.616  &   94.5   & 42.2    \\
            & full finetune & 0.4561  & 0.801& \textbf{72.4} & 28.8\\
            & + multi layer & \textbf{0.5492} & 0.675& 77.8 & \textbf{28.5} \\\midrule
        2000 & baseline &  0.7256  &   0.528  &   108.0   & 56.2    \\
            & full finetune & 0.7274  & 0.602& \textbf{97.5} & 48.7\\
            & + multi layer & \textbf{0.7615} & 0.626 & 100.5 & \textbf{44.7} \\
        \bottomrule
    \end{tabular}
    }
    \vspace*{-1mm}
\end{table}

The TIMIT dataset \cite{timit} was used for evaluation. 
Since each sentence is read by multiple speakers, it can be used for
calculating MTER. Additionally, as the phoneme alignment is manually annotated, 
it also makes PNMI more precise.
We used phonetically compact (SX) set, where each of the 450 sentences
is read by seven different speakers. 
Table \ref{tab:cluster} shows the results for tokens obtained through \textbf{baseline} and \textbf{full finetune}
with the custer sizes of 100 and 2000
from the previous sections.
The results for \textbf{full finetune}
with multiple layers are also presented.

For PNMI, the larger cluster size led to a better result,
which was consistent with the trend in the recognition accuracy shown in Table \ref{tab:single}.
However, among the single-layer cases with the same cluster size,
the values were almost the same regardless of the improvement in recognition accuracy.
When using multiple layers, 
PNMI shows relatively greater improvements compared to recognition accuracies.
These findings indicate that, when the cluster size was fixed, there was no strong consistency between PNMI and recognition accuracy. While SSL fine-tuning contributed more to improving recognition accuracy, the use of multiple layers helped obtain more phoneme-like tokens.
For NQE, \textbf{baseline} achieved the lowest values in all cases.
The large increase in NQE for \textbf{full finetune} is expected due to the absence of 
the k-means loss ($\mathcal{L}^{\text{km}}$) aimed at minimizing quantization error.
This indicates that 
minimizing quantization error is not always required for optimizing downstream tasks.
For TSL and MTER, significant improvements were observed with \textbf{full finetune}. 
The proposed method clearly enhanced 
the robustness to speaker variety and subtle pronunciation differences,
indicating that the tokens have more phonetic properties.

\subsection{Speech resynthesis with acquired tokens}

\begin{table}[tb]
    \centering
    \caption{Resynthesis performance with acquired units:
    synthesis models were trained with LJSpeech and tested with in-domain (ID) and out-of-domain (OOD) data (VCTK).}
    \label{tab:resynth}
    \hspace*{-4mm}
    \vspace*{-5mm}
    \resizebox{1.0\columnwidth}{!}{
    \begin{tabular}{llcccccc}
    \toprule
        cluster&               & MCD & F0 RMSE & \multicolumn{2}{c}{WER [\%]}& \multicolumn{2}{c}{UTMOS}\\
        size        &               & ($\downarrow$)    & ($\downarrow$) & \multicolumn{2}{c}{($\downarrow$)}  & \multicolumn{2}{c}{($\uparrow$)} \\
                    &               & ID    & ID    & ID    & OOD  & ID    & OOD \\\midrule
        100         & baseline      & 6.61  & 0.296 & 7.4   & 31.4   & 3.78  & 3.82  \\
                    & full finetune &  6.68     &  0.297   &  4.0     &    12.4   &   3.78    &    3.76   \\
                    & + multi layer & \textbf{6.53}  &    \textbf{0.290}       &           \textbf{2.7}        &       \textbf{6.1}             &      \textbf{3.85}     &    \textbf{3.89}        \\ \midrule
        2000        & baseline      &    \textbf{6.35}   &   0.292    &    2.7   &  14.0     &   3.86    &    3.43   \\
                    & full finetune &    6.44   &     \textbf{0.282}  &   \textbf{2.6}    &    14.9   &   3.87    &   3.62    \\
                    & + multi layer &  6.40   &  0.285   &   3.2    &      \textbf{7.3} &   3.85    &    \textbf{3.77}   \\\midrule\midrule
        (75)        & g2p           &   7.15    &    0.309   &    3.8   &  3.3     &   3.77    &    3.76   \\
        \bottomrule
    \end{tabular}
    }
\end{table}

We evaluated the performance of speech resynthesis using the tokens
learned through fine-tuning for ASR
to investigate their general applicability to other tasks.
For the synthesis model, we employed Tacotron2 \cite{taco} with the deduplicated token sequences as input.
The training was done using the ESPnet LJSpeech recipe \cite{watanabe18_interspeech, ljspeech},
and for decoding, a Parallel WaveGAN \cite{parallelwavegan} model pretrained 
on LJSpeech\footnote{\texttt{ljspeech\_parallel\_wavegan.v1} available at \url{https://github.com/kan-bayashi/ParallelWaveGAN}}
was used as the vocoder.

The evaluation was conducted from two perspectives:
1) acoustic reconstruction and 2) content reconstruction.
For the former, we used mel-cepstral distortion (MCD) and f0 root mean square error (F0 RMSE) for in-domain (ID) LJSpeech data.
More acoustic information is expected to improve these values.
For the latter, WER by a pre-trained whisper large \cite{whisper} model was used.
This evaluation was 
performed also on out-of-domain (OOD) VCTK corpus \cite{veaux2017cstr} to evaluate the purity of linguistic information in the tokens.
If acoustic information is not discarded and OOD speaker information remains in the tokens, a degradation may be caused 
by the mismatch between the synthesis model trained on LJSpeech data.
UTMOS \cite{saeki22c_interspeech} was also evaluated for overall naturalness.
As a reference, 
we also compared with human-defined phonemes, pure linguistic symbols with no acoustic information, which were
obtained via g2p\footnote{\url{https://github.com/Kyubyong/g2p}}. 

The results are shown in Table\ref{tab:resynth}.
First, when comparing the performance of g2p and discrete token resynthesis, the latter outperformed in many cases except for intelligibility (WER) and UTMOS OOD metrics. 
This suggests that discrete tokens often capture richer acoustic information than phonemes for these metrics.
Among various cluster sizes and training methods, the combination of a cluster size of 100 and \textbf{full finetune} with a multi-layer approach yields the most balanced results across all metrics.
This indicates that, despite ASR being our primary training objective, using a smaller cluster with dedicated fine-tuning leads to a compact and efficient representation for generation purposes.
These findings provide a useful guideline for designing discrete tokens that effectively support both recognition and generation tasks \cite{maiti2024voxtlm,rubenstein2023audiopalm}.

\subsection{Recognition performance with other corpora}
\label{subsec:other}

\begin{table}[tb]
    \centering
    \caption{LibriSpeech-960 (WER [\%]) \& ML-SUPERB (CER [\%]) results with single-layer settings}
    \label{tab:libri960}
    \vspace*{-2mm}
    \resizebox{\columnwidth}{!}{
    \begin{tabular}{llccc}
    \toprule
                   cluster & & LibriSpecch-960 & ML-SUPERB\\
         size & & test-\{clean, other\} & normal/few-shot\\\midrule
        2000 & baseline& 3.0/5.9  & 28.2/41.3  \\
         & full finetune& \textbf{2.8/5.4} &   \textbf{19.9/28.8} \\
        \bottomrule
    \end{tabular}
    }
    \vspace*{-4mm}
\end{table}

Lastly,
we evaluated the recognition performance on LibriSpeech960 
and ML-SUPERB \cite{shi23g_interspeech}.
In both cases, by following the finding in Sections \ref{subsec:libri100} and \ref{subsec:multi}, a single layer with a cluster size of 2000 was used. 
For ML-SUPERB, a multilingual dataset, 
we utilized the 35th layer of XLSR-1B\footnote{\url{https://huggingface.co/facebook/wav2vec2-xls-r-1b}}, following \cite{chang2024exploring}.
For LibriSpeech960, the 21st layer of WavLM-large was used as well.
The results are shown in Table\ref{tab:libri960}.
In both cases, the performance was well improved. 
This shows the validity of our method, even with 
larger or multi-lingual datasets.

\section{Conclusion}
In this study, we introduced differentiable k-means to 
discrete token-based ASR
to enable the joint optimization of tokenization and the downstream task,
along with SSL finetuning and weighted summation across SSL layers.
As a result, ASR performance improved and the obtained tokens exhibited more phonetic characteristics with purer linguistic information. These tokens also showed strong performance even in speech generation tasks.
Future work could extend this approach to other tasks such as speech translation and language modeling.



\newpage

\bibliographystyle{IEEEtran}
\bibliography{mybib}

\end{document}